\documentclass[a4paper]{article}%[a4paper,9pt,francais]
\usepackage[utf8]{inputenc}
\usepackage{amsmath,amsfonts,amssymb,amstext,latexsym}
\usepackage{authblk}

\usepackage{graphicx}

\usepackage{url}

\usepackage[a4paper,left=2cm,right=2cm,top=1.5cm,bottom=1.5cm]{geometry}

\newcommand{\be}{\begin{equation}}
\newcommand{\ee}{\end{equation}}
\newcommand{\bea}{\begin{eqnarray} \nonumber }
\newcommand{\eea}{\end{eqnarray}}
\newcommand{\bi}{\begin{itemize}}
\newcommand{\ei}{\end{itemize}}

\newcommand{\cRm}[1]{\textsc{\romannumeral #1}} % Petit majuscule

\newcommand{\dd}{\mathrm{d}}
\newcommand{\e}{\mathrm{e}}
\newcommand{\ii}{\mathrm{i}}

\newcommand{\coorduo}[2]{% 
    \left\lvert 
      \begin{array}{l}
        #1\\ 
        #2 
      \end{array}  
    \right.% Ne pas oublier le délimiteur invisible. 
  }%}
\newcommand{\hbe}{\hbar_\mathrm{exp}}
\newcommand{\ece}{E_\mathrm{c,\,eff}}
\newcommand{\ve}{v_\mathrm{eff}}
\newcommand{\pe}{p_\mathrm{eff}}
\newcommand{\vecve}{\vec{v}_\mathrm{eff}}
\newcommand{\vecpe}{\vec{p}_\mathrm{eff}}
\newcommand{\vecLe}{\vec{L}_\mathrm{eff}}
\newcommand{\Le}{L_\mathrm{eff}}

\title{Wave-particle duality coming from a bead oscillator in an elastic medium, theoretical study and quantum similarities}
\author{Christian Borghesi\\
\normalsize{Équipe BioPhysStat, Université de Lorraine, 1 boulevard Arago, 57070 Metz, France}}
\date{}

\begin{document}

\maketitle

\abstract{ 
We introduce a dual wave-particle macroscopic system, where a bead oscillator oscillates in an elastic medium which obeys the Klein-Gordon equation. This theoretical system is mostly inspired by bouncing droplets experiments and bead sliding on a vibrating string experiments. This system is studied using a common and simple mathematical formalism. We compute the motion equation of the bead as well as the wave equation of the system. We introduce the effective velocity of the bead with respect to the elastic medium and the wave $\psi$, created by the bead, which modulates the natural wave of the medium. Provided some conditions, $\psi$ obeys an equation analogous to the free Schrödinger equation. In the case of linear and spherical cavities, the particle-like characteristics of the bead, expressed with its effective velocity, are proportional to the corresponding wave-like characteristics of the system.

\scriptsize{{\it This paper is a translation of} Dualité onde-corpuscule formée par une masselotte oscillante dans un milieu élastique : étude théorique et similitudes quantiques.}
}

%%%%%%%%%%%%%%%%%%%%%%%%%%
%%%%%%%%%%%%%%%%%%%%%%%%%%
%%%%%%%%%%%%%%%%%%%%%%%%%%

\section*{Introduction}

For several years, bouncing droplets on a vibrating liquid substrate experiments have shown that macroscopic systems can exhibit behaviours related to wave-particle duality~\cite{yc_walking05}. 
On the one hand, the droplet emits waves in the liquid bath at each bounce and, on the other hand, the waves that the droplet has generated deviate the droplet trajectory. 
Quantum-like phenomena were observed for the first time in classical and macroscopic systems, in which the {\it memory} of the oscillating medium plays a crucial role~\cite{yc_JFM11,yc_chaos14}.
Among other examples, one can cite diffraction and interference with one droplet~\cite{yc_diffraction-interference06}, quantum-like tunneling~\cite{yc_tunnel09}, quantization of orbits~\cite{yc_path-memory-pnas10,yc_SO-eingenstates14,dan_rotating14}, the Zeeman effect~\cite{yc_zeeman12}, or the probability distribution into a cavity~\cite{yc_cavite13}.
See~\cite{bush_review15} for a review. 
Experiments in which the droplet is guided by the waves that it has generated is reminiscent of Louis de Broglie's {\it pilot wave}~\cite{ldb_ondeguidee1927}. 
Discussions on the pilot wave in these experiments are available in~\cite{yc_wavepartduality11,bush_review15,bush_phystoday16}.

However, it seems tricky to formalise mathematically the bouncing droplet problems in order to obtain equations close to the ones of corresponding quantum systems.
This paper aims at presenting a theoretical -- yet practically doable -- system, which mathematical formalism is convenient to identify analogies and differences with quantum systems. 
We extract from the bouncing droplets experiments specific features in order to describe our theoretical system. 
The latter is macroscopic and classical, as it is constituted of a bead oscillator in an elastic medium. 
As will be seen in the following, the system is simple to analyse with the Lagrangian formalism and the covariant formalism based on the d'Alembert equation.
Even if the understanding of this paper does not demand any knowledge of the bouncing droplets experiments, we highly recommend the reading of~\cite{www} and we make several references to these experiments in the text (mainly in footnotes).

The paper is organised as follows. 
In the first section, we formalise an experiment where a bead is free to slide on an oscillating string. 
Many features of this experiment will be used later in the context of our system.
Moreover, this experiment will help visualize the theoretical system that we present in the second section. 
There, we study the dynamics of the bead motion, as well as the wave equations and then focus more precisely on the example of the free bead. 
In the third section, we investigate if an equivalent of the Schrödinger equation governs the wave propagation, without external potential.
Lastly, we consider answering a question {\it à la} de Broglie in the context of our system: can the kinetic of the bead account for the wave characteristics?

%%%%%%%%%%%%%%%%%%%%%%%%%%
%%%%%%%%%%%%%%%%%%%%%%%%%%
%%%%%%%%%%%%%%%%%%%%%%%%%%

\section{Formalisation of the sliding bead on an oscillating string  \label{sexpe}}

Arezki Boudaoud, Yves Couder and Martine Ben Amar have achieved an experiment described in~\cite{yc_sao1999}, where a bead is free to slide on a string -- see \cite{imdelaphys2002} for a less technical presentation. 
The string is harmonically excited by an external source. 
Without any bead, the system is resonant only for specific eigenmodes of the string. 
But a bead slides on the string up to the point where the system becomes resonant for any forcing frequency. 
The authors call this phenomenon self-adaptative behaviour, which is the focus of their study. 
They establish the motion equation of the bead in the neighbourhood of the equilibrium point and the string equation when the bead is at the equilibrium point~\footnote{The authors also study a string with two sliding beads, but this experiment is out of the scope of this work.}. 
In this section, we focus on the formalisation of the experimental system in order to retrieve both the bead equation of motion and the wave equation, and extend their application domain. 
For this purpose, we use a Lagrangian formalism. 
This section makes us familiar with concepts and formalisms that we use in the next section with a more abstract system.\\

%\subsection{Cadre de l'étude de Boudaoud, Couder et Ben Amar \label{ssexpe_cadre}}

The authors of \cite{yc_sao1999} consider the following system:
\bi 
\item[-] A homogeneous string with a linear density $\mu_0$ and a tension $\mathcal{T}$.
Without any external force, the transverse elongation $\varphi(x,t)$ at position $x$ on the string and at time $t$, is governed by the d'Alembert equation with a wave velocity $c_m = \sqrt{\mathcal{T}/\mu_0}$. 
The value of $c_m$ does not have any incidence in this study. 
By contrast, the fact that the waves are transverse is important, as in the case of bouncing droplets.
\item[-] A bead is free to slide on the string. 
The bead is modelled as a punctual mass $m_0$, located on the string at $x=\xi(t)$ at time $t$.
\item[-] Each element of length of the string is excited by an external harmonic force of pulsation $\omega$. 
This excitation is also homogeneous and transverse -- directed along the $(Oz)$ axis. 
The external force acting per element of length is $F_\ell(t) = F_\ell \cos(\omega t)$.
\item[-] Gravity, friction and non-linearities are neglected.
\item[-] The string length is $L$ and its extremities are fixed. 
In this study, the boundary conditions do not play any role since we are only interested in providing the equations of the bead and of the wave.
\ei

%Fig1. Notation
\begin{figure}
\begin{minipage}[c]{0.75\linewidth}
\centering
\includegraphics[width=1.0 \columnwidth,clip=true]{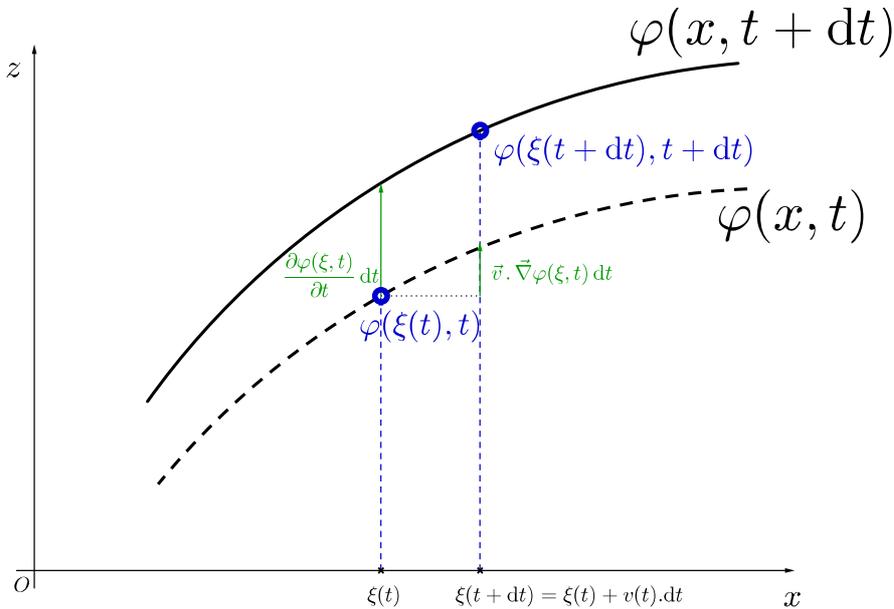}\hfill %width=4cm, height=2.5cm,clip=true
\end{minipage}
\begin{minipage}[c]{0.23\linewidth}
\caption{\small Location of the bead (blue point) on a transversally oscillating string, at times $t$ and $t+\mathrm{d}t$.}
\label{fig1}
\end{minipage}
\end{figure}

%%%%%%%%%%%%%
%%%%%%%%%%%%%
%%%%%%%%%%%%%

\subsection{Formalisation}

The formalism used in this section is inspired by Landau and Lifshitz's electromagnetism formulation (cf. \cite{landau_thchamps}, \S 8, 16 et 27).
The free particle is associated to the Lagrangian $L_0 = -m_0\,c^2\sqrt{1-v^2/c^2}$, which can be approximated by $L_0=1/2\, m_0 v^2$ when $v\ll c$. 
An interaction Lagrangian $L_i = -q(V-\vec{v}.\vec{A})$ is associated to the interaction of the particle with the magnetic field, and the magnetic field has itself a Lagrangian density.

This study is not covariant, in the sense that we do not demand that the bead dynamics exhibit the same invariances as the wave equation. 
In other words, the mechanical waves are governed by the d'Alembert equation and therefore, they are invariant by the Lorentz-Poincaré transformation with the specific celerity $c_m=\sqrt{\mathcal{T}/\mu_0}$, but the dynamics of the bead is not invariant by this transformation.

%%%%%%%%%%%%%
%%%%%%%%%%%%%
%%%%%%%%%%%%%

\subsubsection*{The kinetic energy of the bead on the string described as a sum of a free particle Lagrangian and an interaction Lagrangian}

The bead moves in a 2 dimensional space, which is defined by the axis of the string $(Ox)$, along which the bead motion is usually observed, and the transverse axis of vibration $(Oz)$. 
In the following, this space is called the {\it observable} space, as a reference to one of de Broglie's paper~\cite{ldb_5dim1927}.\\

Let $\vec{r}_{\mathrm{glob}}(t)=\coorduo{\xi(t)}{\varphi\left(x=\xi(t),t\right)}$ be the global location of the bead at time $t$.
Its projection along the observable axis is denoted $\xi(t)$ and its amplitude of vibration is  $\varphi\left(x=\xi(t),t\right)$, see Fig.~\ref{fig1}. 
The velocity vector of the bead, $\vec{v}_{\mathrm{glob}}(t)$ is then
\be %\label{eexpe_vglob}
\vec{v}_{\mathrm{glob}}(t) = \coorduo{\frac{\xi(t+dt)-\xi(t)}{\dd t}}{\frac{\varphi\left(\xi(t+\dd t),t+\dd t \right)-\varphi\left(\xi(t),t \right)}{\dd t}} = \coorduo{v(t)}{\frac{\dd \varphi(\xi,t)}{\dd t} = \frac{\partial \varphi(\xi,t)}{\partial t} + \vec{v}\cdot\vec{\nabla}\varphi(\xi,t)} ,
\ee
where $v(t)=\frac{\dd \xi(t)}{\dd t}$ denotes the {\it observable} velocity of the bead, {\it i.e.} its velocity along the string axis. 
$\frac{\dd \varphi}{\dd t}=\frac{\partial \varphi(\xi,t)}{\partial t} + \vec{v}\cdot\vec{\nabla}\varphi(\xi,t)$, can be understood as the particle velocity of the bead along the transverse axis of vibration. 
Note that the vectors $\vec{v}$ and the gradient of $\varphi$, $\vec{\nabla}\varphi$, are exclusively expressed in the observable space, in contrast to $\vec{v}_{\mathrm{glob}}$.
Here $\vec{v}=v\, \vec{e}_x$ and $\vec{\nabla}\varphi = \frac{\partial \varphi}{\partial x}\vec{e}_x$, where $\vec{e}_x$ denotes the unit vector along $(Ox)$.

The total kinetic energy of the bead is the sum of the observable kinetic energy due to the translation of the bead along the string axis, and the vibrational kinetic energy of the bead. 
The Lagrangian of the bead is its total kinetic energy. 
So the Lagrangian of the bead can be expressed as the sum of $L_0$, the free Lagrangian of the bead along the observable axis,
\be %\label{eexpe_L0}
L_0 = \frac{1}{2}\, m_0\, v^2 \,
\ee
and $L_i$, the interaction Lagrangian between the bead and the field (here the transverse oscillations of the string),
\be \label{eexpe_Li}
L_i = \frac{1}{2}\, m_0\, \left(\frac{\dd \varphi}{\dd t}(\xi,t)\right)^2 \,.
\ee

%%%%%%%%%%%%%
%%%%%%%%%%%%%
%%%%%%%%%%%%%

\subsubsection*{Lagrangian density of the field and Lagrangian density of the external excitation}

The scalar field which is associated to the transverse oscillations, $\varphi(x,t)$, obeys the d'Alembert equation. 
So its Lagrangian density can be written as 
\be \label{eexpe_Lch}
\mathcal{L}_{f} = \frac{1}{2}\, \mathcal{T} \left( \left(\frac{\partial \varphi(x,t)}{c\, \partial t}\right)^2 - \left(\vec{\nabla}\varphi(x,t)\right)^2 \right)\,.
\ee

The Lagrangian density of the external excitation, which is homogeneous and transverse to the string axis, can be expressed as (see {\it e.g.} \cite{landau_meca}, \S 5)
\be 
\mathcal{L}_{e} = F_\ell(t) \,.\, \varphi(x,t) \,.
\ee

%%%%%%%%%%%%%
%%%%%%%%%%%%%
%%%%%%%%%%%%%

\subsection{Equation of motion of the bead}

In~\cite{yc_sao1999}, the observable velocity of the bead is assumed to be small in regards to its vibrational velocity. 
We take advantage of this fact in order to exhibit an amusing analogy with electromagnetism. 
Note that it is expected that this analogy is valid only at low velocities, since electromagnetism is covariant, while the bead dynamics is not.
Just for the beauty of the analogy, we consider either a 2D or 3D observable space, so that the string is either a 2D or 3D elastic medium.

We neglect $v^2$ terms, making the interaction Lagrangian~(\ref{eexpe_Li})
$L_i \approx (m_0/2) \left( \left(\frac{\partial \varphi}{\partial t}\right)^2 + \vec{v}\cdot 2\, \frac{\partial \varphi}{\partial t}\,\vec{\nabla}\varphi \right)$, where the field $\varphi$ is considered at the observable location of the bead $\vec{\xi}$, at time $t$.
The electromagnetic interaction Lagrangian of a particle of charge $q$ in an electric potential $V$ and a magnetic vector potential $\vec{A}$ is $L_i=-\,q\,(V\,-\, \vec{v}\cdot \vec{A})$, see for example \cite{landau_thchamps} \S 16. 
So, the bead dynamics is analogous to the electromagnetic system, when we consider that
\be %\label{eanalogie_em}
\left\{
      \begin{array}{l}
        q \equiv m_0/2\,,\\ 
        V \equiv -\, \left(\frac{\partial \varphi}{\partial t}\right)^2,\\
        \vec{A} \equiv 2\, \frac{\partial \varphi}{\partial t}\,\vec{\nabla}\varphi\,.
      \end{array}  
    \right.
\ee
As in electromagnetism, the equivalent potentials $V$ and $\vec{A}$ only depend on the location $\vec{\xi}$ of the bead at time $t$ \footnote{Without the low velocity approximation $L_i$ is $(m_0/2)\frac{\dd \varphi}{\dd t}(\frac{\partial \varphi}{\partial t} + \vec{v}\cdot\vec{\nabla}\varphi)$, which suggests that $V\equiv -\frac{\dd \varphi}{\dd t} \frac{\partial \varphi}{\partial t}$ and $\vec{A}\equiv \frac{\dd \varphi}{\dd t} \vec{\nabla}\varphi$. 
But then, the equivalent potentials also depend on the bead velocity $\vec{v}$, which is not the case of electromagnetic potentials.}.

When the field is fixed, Euler-Lagrange equation yields the bead equation of motion. 
In the context of the approximation formerly mentioned, an equivalent of the Lorentz force can be seen and the equation of motion is, as one might expect
\be \label{eexpe_mvt}
\frac{\dd }{\dd t}(m_0\,\vec{v}) \approx (m_0/2) \left( -\,\vec{\nabla}V \, -\frac{\partial \vec{A}}{\partial t} \, + \vec{v}\times\vec{\nabla}\vec{A} \right) \,.
\ee

We now consider the specific case of the string, that is to say a 1D observable space, with standing waves.
By taking the temporal average on one oscillation period, the bead only moves under the action of the equivalent scalar potential $V$ -- and so under the potential energy $\langle -\frac{m_0}{2} (\frac{\partial \varphi}{\partial t})^2 \rangle$, where $\langle \cdots \rangle$ denotes the temporal average\footnote{$\frac{\partial \vec{A}}{\partial t}$ is null for standing waves, and $\vec{v}\times\vec{\nabla}\vec{A}$ (which stems from $\vec{\nabla}(\vec{v} \cdot \vec{A}) - (\vec{v}\cdot\vec{\nabla})\vec{A}$) is also null in a 1D observable space.}.
Therefore, we exactly obtain the equation (7) in~\cite{yc_sao1999}.

%%%%%%%%%%%%%
%%%%%%%%%%%%%
%%%%%%%%%%%%%

\subsection{The wave equation}

When the location and the speed of the bead are fixed, the generalized Euler-Lagrange equation yields the wave equation (see Appendix~\ref{sa_expe_ech} for more details):
\be \label{eexpe_ch}
\mathcal{T} \left( \frac{1}{c_m^2} \frac{\partial^2 \varphi}{\partial t^2} - \frac{\partial^2 \varphi}{\partial x^2} \right) = F_\ell(t)\, - \,m_0 \frac{\dd^2 \varphi}{\dd t^2} \cdot \delta\left(x - \xi(t)\right)  \,.
\ee
This equation features the particle acceleration  of the bead, $\frac{\dd^2 \varphi}{\dd t^2}=\frac{\partial}{\partial t}\frac{\dd \varphi}{\dd t} + \vec{v}\cdot\vec{\nabla}\frac{\dd \varphi}{\dd t}$.

It can be applied in a wide variety of contexts, and it appears to be the same as the one observed for a stationary bead (cf. Eq.~(2) in \cite{yc_sao1999})~\footnote{The equation in~\cite{yc_sao1999} features $\frac{\partial^2 \varphi}{\partial t^2}$ rather than $\frac{\dd^2 \varphi}{\dd t^2}$, but both expressions are equivalent when the bead is located at a fixed point $x=\xi$.}.\\

To conclude this section, we have shown that the Lagrangian formalism is suitable for the study of this kind of system, where a bead is free to slide on an oscillating elastic medium.
Equation~(\ref{eexpe_mvt}) shows that the bead moves along the string under the action of the transverse waves on the string. 
Moreover, the wave equation~(\ref{eexpe_ch}) shows that the bead, and more precisely its particle acceleration, is a source of the field. 
This reminds of the bouncing droplets experiments: the droplet is driven by the slope of the transverse wave when it bounces and it also acts as a source of the waves at each bounce.

Besides, with a little exaggeration, the droplet can be understood as a bead which is free to move on the surface of the oscillating bath. 
That encourages us to design a mechanical system where a bead can move on an elastic medium.
This system would be inspired by the bouncing droplets experiments and the one that we have formalised in this section.

%%%%%%%%%%%%%%%%%%%%%%%%%%
%%%%%%%%%%%%%%%%%%%%%%%%%%
%%%%%%%%%%%%%%%%%%%%%%%%%%

\section{The theoretical system proposed and its dynamics \label{sth}}

Let us summarize the main ingredients contained in the bouncing droplets and the sliding bead on a string experiments.
Then, we specify the ingredients that we keep, modify, or reject for the system that we present.  
\bi 
\item[-] The medium supports transverse waves. 
This ingredient is conserved and we also suppose that the waves obey the d'Alembert equation at this step, as in the case of the sliding bead experiment\footnote{This is not the case for the capillary waves of the bouncing droplets experiments.}.
\item[-] The material medium is transversally excited, either by the excitation of the bath itself, as in the bouncing droplets experiments, or by an external Laplace force, as the one exciting the string in the sliding bead experiment. 
It is not a crucial element for our system. 
The essential point in order to discover quantum-like behaviours is that each transverse perturbation at the surface of the bath tends to generate a harmonic oscillation at the Faraday pulsation at the location of the perturbation, as observed in the bouncing droplets experiments (yet not in the sliding bead experiment).
The more this tendency appears, the more can occur quantum-like phenomena~\cite{yc_path-memory-pnas10,yc_SO-eingenstates14}. 
This property of the medium is maintained.
We denote $\Omega_m$, the \textit{natural} pulsation of the standing and transverse waves of the medium in the laboratory frame of reference\footnote{$\Omega_m$ is the analogue of the Faraday pulsation in the bouncing droplets experiments.}.
This property can result from a quadratic potential, that applies to any element of the medium which supports transverse waves.
Thus, due to this property, the transverse waves are not governed by the d'Alembert equation but by the Klein-Gordon equation. 
%
%This tends the waves to oscillate with a precise pulsation.
%
\item[-] In the bouncing droplets experiments (but not in the sliding bead on a string), the elastic medium keeps a {\it memory} of the sustainability of the standing waves. 
We do not use any memory parameter in our system. 
Nevertheless, note that the Klein-Gordon equation naturally exhibits a reminiscence of previous excitations, which decreases with time\footnote{See the term of the Green function of the Klein-Gordon equation, which contains the Heaviside function. Note that this term decreases in time (with the exponent $-3/2$ in 3D), which does not allow an infinite {\it memory}.}.
We discuss below the possibility of a connection between the notion of memory and the parameters of our system.
\item[-] Space is divided between the observable space, where the droplets or the bead move, and an axis along which the transverse vibrations take place. 
This element is also maintained.
\item[-] The bead on a string and the droplet on a bath have no characteristic of their own, except for their mass. 
Here, we make an important modification. 
In our system, the bead is also equipped with an \textit{internal clock}. 
It compels the bead to oscillate transversely with its own pulsation -- denoted $\Omega_0$.
This property originates from a quadratic potential, which belongs to the bead. 
In other words, the bead (or the droplet) described in the previous section is turned into a harmonic oscillator\footnote{This modification is not necessary. 
We could rather choose to stay closer to the bouncing droplets experiments, which implies that there would be no internal clock and $\Omega_0=0$ in all the following equations. 
We have introduced this property because (1) it allows us to express the wave-bead symbiosis condition (cf. section~\ref{sssymbiose}), which directly yields the equivalent Schrödinger equation (cf. section~\ref{ssschrodinger}) and (2) we wanted to directly incorporate -- perhaps wrongly -- in our system a very usual view in quantum mechanics, which states that one may associate a periodic phenomenon of frequency $\nu_0$ to any element of energy with a proper mass $m_0$~\cite{ldb_these}.}.
%
%\og à chaque morceau d'énergie de masse propre $m_0$ soit lié un phénomène périodique de fréquence $\nu_0$ \fg
\ei

Now, we can describe more explicitly and quantitatively the theoretical system that we suggest.

%Fig2. Système théorique
%
\begin{figure}
\begin{minipage}[c]{0.7\linewidth}
\centering
\includegraphics[width=1.0 \columnwidth,clip=true]{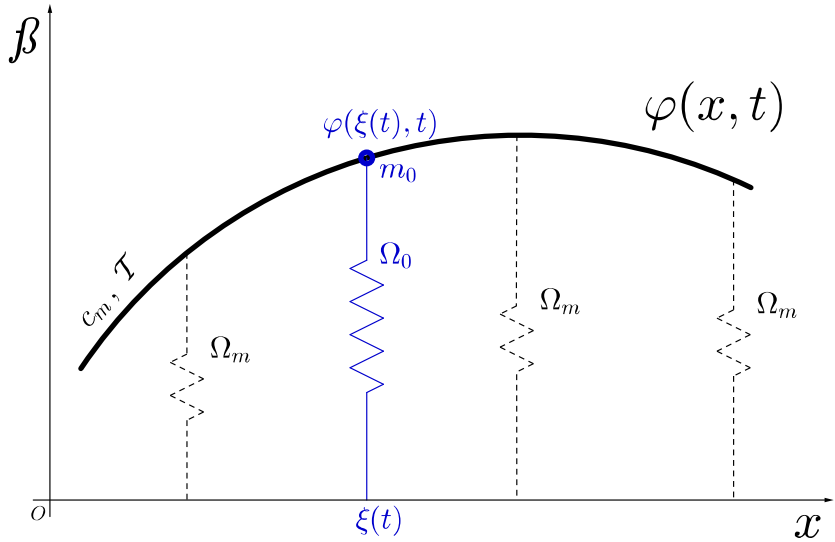}%\hfill %width=4cm, height=2.5cm,clip=true
\end{minipage}
\begin{minipage}[c]{0.28\linewidth}
\caption{\small 
Schematisation of the theoretical system, here in a 1D observable space. 
The quadratic potentials of the medium and of the bead (and their pulsations) are depicted by springs. 
Features of the medium are represented in black, features of the bead are in blue.}
\label{fig2}
\end{minipage}
\end{figure}

%%%%%%%%%%%%%
%%%%%%%%%%%%%
%%%%%%%%%%%%%

\subsection{Framework and formalisation}
The system (see Fig.~\ref{fig2} for a schematic representation) is constituted as follows:
\bi 
\item[-] An homogeneous, isotropic, elastic medium of one, two or three dimensions (where $\vec{r}$ denotes its coordinates, $(x)$, $(x,y)$ or $(x,y,z)$). 
It is subjected to a \textit{tension} $\mathcal{T}$ and has a mass per element of length/surface/volume $\mu_0$. 
Without any external source, and without considering the tendency of the medium to support standing waves at pulsation $\Omega_m$ (see below), it supports transverse waves governed by the d'Alembert equation, where the velocity of the waves is $c_m=\sqrt{\mathcal{T}/\mu_0}$. 
Here again, the value of $c_m$ is not important. 
Note that the medium is neither dispersive nor dissipative.
\item[-] The transverse wave at point $\vec{r}$ and time $t$ has an amplitude $\varphi(\vec{r},t)$ and a direction along the {\it eszett} axis, $(O\ss)$, to give it a name.
\item[-] The elastic medium has a potential per element of length/surface/volume $\frac{\mu_0}{2}\,\Omega_m^2\,\varphi^2(\vec{r},t)$. 
It forces the medium to support a transverse standing wave at pulsation $\Omega_m$ in the laboratory reference frame. 
Due to this feature, $\varphi(\vec{r},t)$ is a scalar field which obeys the Klein-Gordon equation -- when there is neither any source nor external excitation.
\item[-] A bead is considered as a punctual mass, $m_0$, located on the elastic medium at $\vec{r}=\vec{\xi}$ and time $t$. 
The bead is free to slide on the medium.
\item[-] The bead is now a harmonic oscillator.
Its potential is $\frac{m_0}{2}\,\Omega_0^2\,\varphi^2(\vec{\xi},t)$. 
It implies that the bead oscillates transversely at its natural pulsation $\Omega_0$.
\item[-] The space is separated between the observable space, in which it is simple to explicit for example the particle velocity of the bead $\vec{v}=\frac{\dd \vec{\xi}}{\dd t}$, or the slope of the field $\vec{\nabla}\varphi$, and an axis which is specific to the vibrations. 
In the following, we do not use any longer the term {\it observable} when referring to the location, the speed, etc. of the bead.
\item[-] Gravity, friction, non-linearities and other dissipative effects are neglected in theory. 
\ei 
In this section, the study is covariant with respect to the elastic medium and in particular to the speed of the waves $c_m$. 
As the d'Alembert or the Klein-Gordon equation must be satisfied in all our reference frames -- in uniform translation between themselves -- we must use the Lorentz-Poincaré transformation (with $c_m$). 
This transformation leaves indeed the d'Alembert or Klein-Gordon equation invariant.
% which applies to the field $\varphi$ 
%
In other words, we do as if the Lorentz-Poincaré transformation (with $c_m$) can be applied to our system, because we want the d'Alembert (or Klein-Gordon) equation to govern the field $\varphi$ in any reference frame. 
This is a major modification in regards to the sliding bead experiment, and even more in regards to the current formalisations of the bouncing droplets experiments.

In our system, we assume that the usual relativistic effects are negligible, {\it i.e.} either $c_m\ll c$ or $c_m=c$, where $c$ is the speed of light in a vacuum. 
In the following, it is assumed that the system is covariant and that the Lorentz-Poincaré transformation using $c_m$ concerns the elastic medium.
We also neglect the behaviours of the bead which do not stem from its interaction with the elastic medium.
It implies that the vibrational speed of the bead is much greater than its sliding velocity in the laboratory reference frame. 
In short, we only consider the interaction of the bead with the elastic medium, as if nothing else was interacting with it.

Let us now express the Lagrangian densities of this system. 
Calculations are made in a 3D space, however, they would be made experimentally in 1D or 2D.
We also use the usual covariant formalism.\\

The total Lagrangian density of the bead comes from its kinetic energy of vibration and its natural potential energy. 
That is, in a covariant form:
\be \label{eth_lm}
\mathcal{L}_m = \frac{1}{2}\,\rho_0 \left[\left(\frac{\dd \varphi}{\dd \tau}\right)^2 \; -\; \Omega_0^2\,\varphi^2 \right]\,,
\ee
where $\tau$ and $\rho_0$ respectively denote the natural time of the bead and its natural mass density. 
$\rho_0=m_0\,\delta(\vec{r}_0)$ when it is expressed in the reference frame of the bead itself, denoted $\mathcal{R}_0$, in which the bead is assumed to be at the origin.

The Lagrangian of the bead, $L_m=\int \mathcal{L}_m \dd^3\vec{r}$, expressed in the reference frame $\mathcal{R}$ where the bead has a velocity $\vec{v}$ is
\be \label{eth_Lm}
L_m = \frac{m_0}{2\gamma} \left[\left(\frac{\dd \varphi}{\dd \tau}\right)^2 \,-\,\Omega_0^2\,\varphi^2\right] \,,
\ee
where $\gamma=\frac{1}{\sqrt{1-v^2/c_m^2}}$.

The Lagrangian density of the field includes a part due to the d'Alembert equation, and another part due to the tendency of the medium to  support standing waves at pulsation $\Omega_m$:
\be \label{eth_lch}
\mathcal{L}_{f} = \frac{1}{2}\, \mathcal{T} \left[ \partial_\mu \varphi \, \partial^\mu \varphi \, - \, \frac{\Omega_m^2}{c_m^2} \varphi^2 \right] \,.
\ee

%%%%%%%%%%%%%
%%%%%%%%%%%%%
%%%%%%%%%%%%%

\subsection{The equation of motion of the bead and the wave equation}

The guidance equation of the bead can be derived from $L_m$ (see Appendix~\ref{sa_th_emvt} for more details) and can be written as:
\be \label{eth_mvt_noncov}
\frac{\dd}{\dd \tau}\left[\gamma\, \frac{m_0}{2\,c_m^2}\left(\left(\frac{\dd \varphi}{\dd \tau}\right)^2 +\, \Omega_0^2\,\varphi^2\right)\vec{v}\right] = -\, m_0 \,\left(\frac{\dd^2 \varphi}{\dd \tau^2} + \Omega_0^2\,\varphi \right)   \, \vec{\nabla}\varphi  \,.
\ee
Using a covariant formulation\footnote{Eq.~(\ref{eth_mvt}) can also be expressed as $\frac{m_0}{2\,c_m^2}\left[\left(\frac{\dd \varphi}{\dd \tau}\right)^2 + \Omega_0^2\,\varphi^2\right] \,\frac{\dd U_\mu}{\dd \tau} = m_0 \,\left(\frac{\dd^2 \varphi}{\dd \tau^2} + \Omega_0^2\,\varphi \right)   \, \left(\partial_\mu \varphi - \frac{U_\mu}{c_m^2} \frac{\dd \varphi}{\dd \tau}\right)$.}:
\be \label{eth_mvt}
\frac{\dd}{\dd \tau}\left[\frac{m_0}{2\,c_m^2}\left(\left(\frac{\dd \varphi}{\dd \tau}\right)^2 + \Omega_0^2\,\varphi^2\right)U_\mu\right] = m_0 \,\left(\frac{\dd^2 \varphi}{\dd \tau^2} + \Omega_0^2\,\varphi \right)   \, \partial_\mu \varphi  \,,
\ee
where $U_\mu$ denotes a covariant component of the 4-velocity of the bead, $U^\mu=\frac{\dd \xi^\mu}{\dd \tau}$, and $\partial_\mu \varphi = \frac{\partial \varphi}{\partial x^\mu}$ denotes the 4-gradient of $\varphi$. 
The values of $\varphi$ and of its gradient are considered at the location of the bead.
This equation shows that the bead is deflected by the field, in particular by its gradient at the position of the bead. 
Note that the deflection of the bead does not depend on its mass, as for a system submitted to a gravitational force.

The wave equation also stems from the principle of least action (see Appendix the Appendix~\ref{sa_th_ech} for more details) and is written as follows:
\be \label{eth_ch}
\frac{1}{c_m^2}\frac{\partial^2 \varphi}{\partial t^2} \,-\, \Delta\varphi \;+\; \frac{\Omega_m^2}{c_m^2}\,\varphi = -\,\frac{\rho_0}{\mathcal{T}}\,\left(\frac{\dd^2 \varphi}{\dd \tau^2} + \Omega_0^2\,\varphi \right) \,,
\ee
where $\Delta$ denotes the Laplacian operator. 
This equation can be interpreted as an inhomogeneous Klein-Gordon-like equation, in the presence of a field source.
The source is localised at the position of the bead and depends on the vibration of the bead.

We briefly comment on these two coupled equations. 
On the one hand, it clearly appears that the bead is the source of the wave and on the other hand, the bead is piloted and guided by the waves that it has generated. 
Thus, the bead oscillator which is free to slide on an elastic medium is a dual system, like the bouncing droplets previously mentioned. 
In the following, we show that the wave-particle duality is also quantitatively expressed, when the characteristics of the wave are compared to the characteristics of the bead. 
We even observe that there is a better correspondence in this case than in the case of the bouncing droplets experiments~\cite{wave-mediated}.

%%%%%%%%%%%%%
%%%%%%%%%%%%%
%%%%%%%%%%%%%

\subsection{Symbiosis between the bead and the wave \label{sssymbiose}}

It is now interesting to investigate the special case -- which has substantial consequences -- where the bead is not the wave source any longer, {\it i.e.}
\be \label{eth_symbiose}
\left(\frac{\dd^2 \varphi}{\dd \tau^2} + \Omega_0^2\,\varphi \right) = 0 \,.
\ee
Thus, the bead oscillates in its reference frame at its natural pulsation $\Omega_0$. 
The amplitude of the vibration of the bead $\varphi_M$ in $\mathcal{R}_0$ is then $\varphi_M(\tau) = A_M\,\cos(\Omega_0\,\tau)$, where $A_M$ is its maximum.

Note that the previous relation implies that the field does not exhibit a singularity located at the position of the bead.
In other words, its amplitude does not diverge.

When the previous relation is satisfied, $[(\frac{\dd \varphi}{\dd \tau})^2 + \Omega_0^2\,\varphi^2]$ remains constant\footnote{Indeed, $\frac{\dd}{\dd \tau}[(\frac{\dd \varphi}{\dd \tau})^2 + \Omega_0^2\,\varphi^2]=2\,\frac{\dd \varphi}{\dd \tau}(\frac{\dd^2 \varphi}{\dd \tau^2} + \Omega_0^2\,\varphi)$.}.
According to Eq.~(\ref{eth_mvt}), the bead is not deflected by the field. 
The bead either remains at rest or it is in uniform motion, {\it i.e} $\frac{\dd U_\mu}{\dd \tau} = 0$, the opposite would have been surprising for a dual wave-particle system.

It is tempting to define the {\it natural energy}, $E_0$, of the bead as
\be \label {eth_E0}
E_0=\frac{m_0}{2}\left[\left(\frac{\dd \varphi}{\dd \tau}\right)^2 + \, \Omega_0^2\,\varphi^2\right] \,,
\ee
that is the sum of its kinetic and potential energies in covariant formulation. 
Thus, the bead natural energy remains constant when the relation~(\ref{eth_symbiose}) is satisfied. 
This point can be related to the fact that $\varphi_M(\tau) = A_M\,\cos(\Omega_0\,\tau)$, {\i.e.} the bead harmonically oscillates at its natural pulsation in $\mathcal{R}_0$. 
We could be tempted to set the experimental values in such a way that $A_m\,\Omega_0=\sqrt{2}\,c_m$, {\it i.e.} $E_0=m_0\,c_m^2$.
However, we will see in the following that the value of $E_0$ does not matter when looking for quantum analogies.\\

In summary, after the bead has ``filled'' the elastic medium with its mark sufficiently for the relation (\ref{eth_symbiose}) to be satisfied -- {\it i.e.} the bead oscillates harmonically at pulsation $\Omega_0$ in $\mathcal{R}_0$ -- the maximum amplitude and the natural energy of the bead remain constant in time. 
Then, the bead is neither the source of the waves, nor is it deflected by the waves\footnote{Which simplifies computations$\ldots$}. 
In other words, everything is as if the bead does not have any effect on the waves and the waves do not act on the bead, so that an observer who focuses on the waves would be blind to the bead. 
However, these two phenomena -- the wave and the motion of the bead -- are intimately connected. 
When the relation (\ref{eth_symbiose}) is satisfied, the wave and the bead are said to be {\it in symbiosis}.

This state of symbiosis means that there is a strong relationship, or an intimate harmony, between the wave and the bead. 
Therefore, we will not be surprised, when there is symbiosis, to see that the wave state is reflected in the bead kinetics -- and vice versa.
In other words, the symbiosis state allows to quantitatively express the wave-particle duality. 
We will do so in the following section, when the wave description, as it is commonly formulated in quantum mechanics handbooks, is associated with the bead kinetic.

But first, it is appropriate to illustrate what we have been described here by a simple example (that we use later), that is a free bead in symbiosis.

%%%%%%%%%%%%%
%%%%%%%%%%%%%
%%%%%%%%%%%%%

\subsection{Free bead in symbiosis with a wave}

Consider a bead in symbiosis with the field. 
First, we are looking for conditions (between pulsations $\Omega_m$ and $\Omega_0$) that allow the symbiosis, then we give the expression of the field.

In the natural reference frame $\mathcal{R}_0$, the symbiosis condition -- see~(\ref{eth_symbiose}) -- implies that the bead oscillates at pulsation $\Omega_0$, for example at the origin of this frame. 
We assume that the wave is centred on the bead, spherically symmetric and stationary in $\mathcal{R}_0$. 
Thus, the wave field in $\mathcal{R}_0$ -- which associated time $t_0$ is also the proper time of the bead -- writes as $\varphi=F(r_0)\,\cos(\Omega_0\,t_0)$. 
Using this expression in equation (\ref{eth_ch}), here without any wave source, it comes that $\Delta F = -\frac{\Omega_0^2 - \Omega_m^2}{c_m^2}F$. 
In order for the field not to be divergent neither at the origin nor at infinity, it is necessary that $\Omega_0 > \Omega_m$~\footnote{The solution for $\Omega_0 = \Omega_m$ is in $1/r_0$, while the solution for $\Omega_0 < \Omega_m$ is in $\cosh(\sqrt{\Omega_m^2 - \Omega_0^2}\,r_0/c_m)/r_0$ and/or $\sinh(\sqrt{\Omega_m^2 - \Omega_0^2}\,r_0/c_m)/r_0$.}.
In other words, having $\Omega_0 > \Omega_m$ is the condition that allows a free bead to be in symbiosis with a field -- stationary and with spherical symmetry.
Thus, denoting $A_M$ the maximum amplitude at the origin, the wave field in $\mathcal{R}_0$ can be written
\be \label{eth_varphiR0}
\varphi(\vec{r_0},\,t_0) = A_M \, \frac{\sin(K_0\,r_0)}{K_0\,r_0}\:\cos(\Omega_0\,t_0)\,,~\mathrm{with}~K_0^2 = \frac{\Omega_0^2 - \Omega_m^2}{c_m^2}\,. 
\ee
In 2 dimensions, $\Delta F = -K_0^2\,F$ yields $F(r_0) = A_M\, \mathrm{J}_0(K_0\,r_0)$, where $\mathrm{J}_0$ denotes the Bessel function of the first kind of order $0$.

Let us write the expression of the wave in the laboratory reference frame $\mathcal{R}$, where the bead has a velocity $v$, directed for example along $(Ox)$.
Thus, $\mathcal{R}_0$ has a velocity $\vec{v}$ along the $(Ox)$ axis of $\mathcal{R}$.
The Lorentz-Poincaré transformation in (\ref{eth_varphiR0}) yields
\be \label{eth_varphiR}
\varphi(\vec{r},\,t) = F\left(\sqrt{\gamma^2\,(x-v\,t)^2 + y^2 + z^2}\right) \; \cos(\Omega\,t - k\,x) \,,
\ee
where $\gamma=\frac{1}{\sqrt{1-v^2/c_m^2}}$, $\Omega = \gamma\,\Omega_0$, $k= \gamma\,\Omega_0\,v/c_m^2$ and $F(X)=A_M\,\frac{\sin(K_0\,X)}{K_0\,X}$. 
At this step, there is no direct connection between $k$ and $K_0$.
The phase of this travelling wave is plane and its amplitude propagates at the velocity of the bead~\footnote{The main difference between this wave and the wave at the surface of the bath in the bouncing droplets experiments, where $F(X)=A_M \, \mathrm{J}_0(K_0\,X)$ in 2D, is the travelling nature of the wave.
For the bouncing droplets, the wave is standing and its phase term writes as $\cos (\Omega\,t)$.}.
This wave is very similar to the Barut wave, very nicely used in quantum mechanics by Laurent Bindel~\cite{lbindel_mq12,lbindel_mqxx}, where the phase term is the de Broglie phase.
It notably features the de Broglie wavelength.

In the laboratory reference frame, the bead is located on $x=v\,t$ (and $y=z=0$).
Thus, it comes from equation~(\ref{eth_varphiR}) that the vibration of the bead follows $\varphi_M(t)=\varphi(x=v\,t,\,y=0,\,z=0,\,t) = A_m\,\cos(\frac{\Omega_0}{\gamma}t)$. 
The bead harmonically oscillates in $\mathcal{R}$ at pulsation $\Omega_0/\gamma$. 
According to the bouncing droplets experiments, it is natural to consider that the bead oscillates harmonically at pulsation $\Omega_m$ in $\mathcal{R}$~\footnote{Nevertheless, in the bouncing droplets experiments, both the standing wave and the droplets oscillate at the Faraday pulsation in $\mathcal{R}$, which is the pulsation analogue to $\Omega_m$. 
Here, only the bead oscillates at pulsation $\Omega_m$.}.
It follows a relation between the two pulsations of the system ($\Omega_0$ and $\Omega_m$) and the velocity of the bead in the laboratory reference frame, $\mathcal{R}$, that is $\gamma = \Omega_0/\Omega_m$. 
The greater the ratio  $\Omega_0/\Omega_m$, the greater is the velocity of the bead in $\mathcal{R}$~\footnote{In the bouncing droplets experiments, the increase of the velocity of the bead depends on the memory~\cite{yc_JFM11}. 
This suggests a possible relation between the memory in the bouncing droplets experiments and $\Omega_0-\Omega_m$ here. 
This could be an interesting matter to investigate, but it is out of the scope of this paper.}.

To summarize, we have seen that (1) the symbiosis condition between the bead and the wave implies that $\Omega_0 > \Omega_m$, (2) the wave has a progressive and plane phase with a travelling and localised amplitude, (3) the bead oscillates in the laboratory reference frame at pulsation $\Omega_m$ and (4) the greater $\Omega_0$ is, the greater is the velocity of the bead.

Lastly, it is interesting to describe the system when the velocity of the bead is small in regards to $c_m$, {\it i.e.} $\Omega_0$ is slightly greater than $\Omega_m$. 
We show in the next section that this case satisfies a Schrödinger-like equation.
But first, we need to carry out some calculations, which, although not fascinating, will be needed later on.

%%%%%%%%%%%%%
%%%%%%%%%%%%%
%%%%%%%%%%%%%

\subsubsection*{Low velocity approximation, {\it i.e.} $\Omega_m \lesssim \Omega_0$}

In this section, we neglect the terms of higher order than $v^2/c_m^2$.
Pulsation $\Omega$ of $\varphi$ in $\mathcal{R}$ (cf. Eq.~(\ref{eth_varphiR})) is slightly greater than $\Omega_m$ by a pulsation $\omega$, such that
\be \label{eth_omega}
\omega \approx\Omega_m\,\frac{v^2}{c_m^2} \,,
\ee
because $\Omega=\Omega_m+\omega$ and $\Omega = \gamma \, \Omega_0 = \gamma^2 \, \Omega_m$. 
Moreover, following this approximation $K_0$ and $k$ are equal, {\it i.e.} $K_0\approx k \approx \Omega_m\, v/c_m^2$.

The wave field (cf. Eq.~(\ref{eth_varphiR})) writes as $\varphi = F \; \cos((\Omega_m + \omega)\,t - k\,x)$, where $\Delta F \approx -K_0^2\,F$, {\it i.e.} $F(X)$ has the same expression as the one above but here $F \approx F(\sqrt{(x-v\,t)^2 + y^2 + z^2})$.

Let us write the {\it local wave-vector}, $\vec{\kappa}$, associated to $\varphi$ at the location of the bead $(x=v\,t,\,y=z=0)$.
Since $\Delta F \approx -K_0^2\,F$, the above expression of $\varphi$ in the wave equation (\ref{eth_ch}) without source term yields $-(\Omega_m+\omega)^2/c_m^2 + k^2 + K_0^2 + \Omega_m^2/c_m^2 \approx 0$. 
Now, $\varphi \propto \cos\left((\Omega_m + \omega)\,t - \kappa\,x\right)$ (at the location of the bead) in the homogeneous wave equation (\ref{eth_ch}) yields $-(\Omega_m+\omega)^2/c_m^2 + \kappa^2 + \Omega_m^2/c_m^2 = 0$. 
By identification, it comes that
\be \label{eth_kappa}
\vec{\kappa} \approx \sqrt{k^2 + K_0^2}\,\vec{e}_x \approx \sqrt{2}\:\Omega_m\, \frac{v}{c_m^2}\,\vec{e}_x \,.
\ee
And the relation between $\omega$ and $\kappa$ follows
\be \label{eth_omega_kappa2}
\omega \approx \frac{\kappa^2\,c_m^2}{2\;\Omega_m} \,. 
\ee

We can now investigate the wave-particle duality of this theoretical system.
In particular, we consider the correspondence between the characteristics of the wave and of the particle in order to draw quantitative consequences.

%%%%%%%%%%%%%
%%%%%%%%%%%%%
%%%%%%%%%%%%%

\subsection{Agreement between the wave and the particle characteristics, effective velocity and modulating wave}
{\scriptsize {\it This section changes the order followed in the French version in order to make it shorter. Contrarily to the French version, we directly define the effective velocity and the modulating wave. The numbering of the equations follows the one of the French version.}}

In this section, we adopt the elastic medium point of view and we are looking for the effects of the bead on it. 
We ask the following question: How does the elastic medium ``perceive'' the presence of the bead?
In other words, according to the elastic medium point of view, how can we describe and quantify the changes due to the bead? 
In this section, we consider that the low velocity approximation is satisfied, {\it i.e.} $\Omega_0$ is slightly greater than $\Omega_m$, and from now on, we use the notation $=$ instead of $\approx$.

Let us recall that the elastic medium tends to support standing waves at pulsation $\Omega_m$ in the laboratory reference frame $\mathcal{R}$. 
This plays the role of the reference behaviour of the elastic medium without the bead. 
The presence of the bead in the elastic medium is thus described by comparison to this ``normal'' behaviour of the medium.
This is the crucial point of our reasoning.
The two aspects of the presence of the bead in the elastic medium (the one of the waves and the one of the particle) must be described with regards to this reference behaviour. 
It leads to the definition of the {\it effective velocity} $\vecve$, of the bead on the one hand, and of the {\it modulating wave} $\psi$, on the other hand.

The total wave $\varphi$ corresponding to the free bead in the elastic medium is given by Eq.~(\ref{eth_varphiR}), where $\Omega=\Omega_m+\omega$. 
This wave can be interpreted as the modulation of the \textit{natural wave} of the elastic medium by a specific wave. 
This example shows that the bead does not generate an additional wave to the natural wave of the elastic medium, but modulates the natural wave. 
We call $\psi$ the modulating wave, which is specifically associated to the bead. 
It is convenient to use the complex notation~\footnote{To convince us, consider the term $\cos((\Omega_m+\omega)\,t - k\,x)$ in  Eq.~(\ref{eth_varphiR}).} of $\varphi$, which writes now as
\be \label{eth_psi_def}
\tag{25}
\varphi(\vec{r},\,t) = \mathrm{Re}\left[ \psi(\vec{r},\,t) ~ \e^{-\ii\,\Omega_m\,t}  \right] \,,
\ee
where $\mathrm{Re}$ denotes the real part. 
For example, in the low velocity approximation, $\psi(\vec{r},\,t) = F(\vec{r},\,t) \, \e^{\ii(k\,x\,-\,\omega\,t)}$, where $F$ has the same form as above. 
Using the local wave-vector, $\psi(\vec{r},\,t) \propto \e^{\ii(\vec{\kappa}\cdot\vec{r}\,-\,\omega\,t)}$.

We define the velocity at which the elastic medium (in its normal regime) perceives the bead. 
Without the bead, the elastic medium tends to support standing waves at pulsation $\Omega_m$ in $\mathcal{R}$. 
The phase term of $\varphi$ would then be $(\Omega_m\,t)$. 
But in the presence of the bead, the wave is altered. 
It is convenient to use the local form of $\varphi$ at the location of the bead. 
This form has indeed only one progressive term in the phase, while the form in Eq.~(\ref{eth_varphiR}) has two, one in the phase and another in the amplitude. 
The phase term can be written $((\Omega_m+\omega)\,t \, - \, \vec{\kappa}\cdot\vec{r}\,\,)$. 
This modification of the phase is due to the presence of the bead. 
From the elastic medium point of view (with the reference pulsation $\Omega_m$), the wave behaves as if its phase were changed by a Lorentz-Poincaré transformation with velocity $\vecve$. 
According to this point of view, the phase term writes as $(\gamma_{\mathrm{eff}}\,\Omega_m\,t \,-\,\gamma_{\mathrm{eff}}\,\frac{\Omega_m}{c_m^2}\vecve\cdot\vec{r})$, where $\gamma_{\mathrm{eff}} = 1/\sqrt{1-\frac{\ve^2}{c_m^2}}$. 
By identification, the presence of the bead is associated to an effective velocity, $\vecve$, which is perceived by the elastic medium.
In the low velocity approximation, $\vecve$ is such that
\be \label{eth_ve}
\tag{26}
\omega = \Omega_m\frac{\ve^2}{2\,c_m^2} \hspace*{1cm}\mathrm{and}\hspace*{1cm} \vec{\kappa}=\frac{\Omega_m}{c_m^2}\,\vecve\,.
\ee

It is easy to evaluate the effective velocity of the free bead. 
From Eqs.~(\ref{eth_omega}) and (\ref{eth_kappa}) and the definition of the effective velocity (\ref{eth_ve}), it comes that $\vecve = \sqrt{2}\:v\,\vec{e}_x$.
The question that we want to answer now is: Why is the effective velocity different from the velocity $v$ of the bead? 
To answer this, remember that the velocity $v$ is related to the velocity between the reference frames $\mathcal{R}$ and $\mathcal{R}_0$.
We have seen that the pulsation $\Omega$ of $\varphi$ is such that $\Omega=\Omega_0/\sqrt{1-\frac{v^2}{c_m^2}}$. 
But from the point of view of the elastic medium, the reference frame $\mathcal{R}_0$ as well as the pulsation $\Omega_0$ of the bead have no meaning.
Thus, the velocity $v$ has no meaning either from the elastic medium point of view, in contrast to the effective velocity.
As a consequence, the effective velocity can be non-zero, corresponding to measurable effects on the elastic medium, while the bead seems stationary for an observer in $\mathcal{R}$.\\

The effective velocity allows us to connect wave characteristics to particle characteristics.
Let us define the effective kinetic energy and the effective momentum respectively as $\ece = \frac{1}{2}\,m_0\,\ve^2$ and $\vecpe=m_0\,\vecve$. 
Then, from Eq.~(\ref{eth_ve}) it is convenient to define
\be \label{eth_hbe}
\tag{22}
\hbe = \frac{m_0\,c_m^2}{\Omega_m} \,.
\ee
Using $\hbe$, the effective kinetic energy writes as
\be \label{eth_ece_omega}
\tag{23}
\ece = \hbe \, \omega %\,,
\ee
and the effective momentum as
\be \label{eth_pe_kappa}
\tag{24}
\vecpe = \hbe \; \vec{\kappa} \,.
\ee
For the free bead in the case of the low speed approximation, the relation $\ece = \frac{\pe^2}{2\,m_0}$, is consistent with both Eq.~(\ref{eth_omega_kappa2}) and the definition of $\hbe$ (\ref{eth_hbe}).

$\hbe$ appears as a proportionality coefficient between wave characteristics and particle characteristics. 
$\hbe$ is neither an equivalent of the reduced Planck's constant nor a constant proper to the elastic medium, as it depends not only on characteristics of the medium (through $\Omega_m$ and $c_m$) but also on the mass of the bead $m_0$. 
This mass is arbitrary in our thought-experiment. 
There is no relation between these three values. 
So, each bead with its own mass implies its own proportionality coefficient $\hbe$. 
Nevertheless, with regards to the system that we study, $\hbe$ as defined by (\ref{eth_hbe}), is the proportionality coefficient between waves characteristics ($\omega$ and $\vec{\kappa}$) and particle characteristics (kinetic energy and momentum, both being effective).\\

To sum up, we have seen that the presence of the bead in the elastic medium implies a modulating wave, $\psi$ (cf. Eq.~(\ref{eth_psi_def})).
It modulates the natural standing wave (at pulsation $\Omega_m$) of the elastic medium in $\mathcal{R}$. 
The characteristics of the wave due to the presence of the bead in the elastic medium -- the additional pulsation $\omega$ and the local wave-vector $\vec{\kappa}$ at the location of the bead, both contained in $\psi$ -- are proportional to the effective characteristics of the kinetics of the bead (its kinetic energy (\ref{eth_ece_omega}) and its momentum (\ref{eth_pe_kappa})). 
The effective velocity of the bead is defined (cf. Eq.~(\ref{eth_ve})) via the reference pulsation $\Omega_m$ of the elastic medium. 
The proportionality coefficient between the characteristics of the wave and the characteristics of the particle, $\hbe$ (cf. Eq.~(\ref{eth_hbe})), depends on the parameters of the elastic medium and also on the mass of the bead $m_0$.

This section concerned a free bead in symbiosis with the wave. 
It is now interesting to turn to other systems in order to know whether the proportionality between the wave and the particle characteristics stands.

%%%%%%%%%%%%%%%%%%%%%%%%%%
%%%%%%%%%%%%%%%%%%%%%%%%%%
%%%%%%%%%%%%%%%%%%%%%%%%%%

\section{Quantum similarities \label{ssim}}

\setcounter{equation}{26}

In this section, we consider that (1) the bead and the wave are in symbiosis, (2) the system is in a stationary state, (3) the velocity of the bead is much lower than $c_m$ and (4) there is no external potential acting on the bead and/or the wave field. 
We deal with a bead in a cavity, this allows us to answer a question in the line of de Broglie's (see {\it e.g.} \cite{ldb_tentative}, \S \cRm{11}) in this specific case: Does the kinetic of the bead explain the state of the system? 
Or in other words, is the stationary wave in the cavity reflected in the kinetic of the bead, and {\it vice versa}?

But first of all, we should establish the wave equation which governs the modulating wave $\psi$ when the bead and the wave are in symbiosis.

%%%%%%%%%%%%%
%%%%%%%%%%%%%
%%%%%%%%%%%%%

\subsection{The equivalent Schrödinger equation for the modulating wave \label{ssschrodinger}}

$\varphi$ is governed by the wave equation (\ref{eth_ch}) without source term, because the bead and the wave are in symbiosis. 
Using the modulating wave $\psi$ contained in $\varphi$ (\ref{eth_psi_def}) and after elimination of $\e^{-\ii \Omega_m t}$, it comes that $-2\,\ii\frac{\Omega_m}{c_m^2}\frac{\partial \psi}{\partial t} - \frac{1}{c_m^2}\frac{\partial^2 \psi}{\partial t^2} - \Delta\psi = 0$. 
In the low velocity approximation, the term $\frac{\partial^2 \psi}{\partial t^2}$ is negligible with respect to the other terms\footnote{The calculations are similar to those in the literature, when the Schrödinger equation is deduced from the Klein-Gordon equation. 
But to convince ourselves of the validity of this approximation, we express the wave $\varphi$ of the free bead with the local wave-vector.
According to Eqs.~(\ref{eth_omega}) and (\ref{eth_kappa}), the term $\frac{1}{c_m^2}\frac{\partial^2 \psi}{\partial t^2}$ is proportional to $\frac{\Omega_m^2}{c_m^2}\frac{v^4}{c^4}$, while the two other terms are proportional to $\frac{\Omega_m^2}{c_m^2}\frac{v^2}{c^2}$.}. 
Thus, the modulating wave associated to the presence of the bead in the elastic medium is governed by the equation
\be \label{esim_schroeq}
\ii \,\frac{\partial \psi}{\partial t} = -\,\frac{c_m^2}{2\,\Omega_m}\,\Delta\psi\,.
\ee
The previous equation not only resembles the free Schrödinger equation, it also has the exact same form when using the proportionality coefficient defined in (\ref{eth_hbe}):
\be \label{esim_schro}
\ii \,\frac{\partial \psi}{\partial t} = -\,\frac{\hbe}{2\,m_0}\,\Delta\psi \,.
\ee

Remember that the total wave $\varphi$ obeys a Klein-Gordon like equation without source term (\ref{eth_ch}), while the modulating wave -- which expresses pointedly the presence of the bead in the elastic medium -- is governed by the equivalent Schrödinger equation (\ref{esim_schroeq}). 
The latter is valid when the bead has a low velocity with respect to $c_m$ and when it is in symbiosis or more generally, when the bead is not a source of the wave\footnote{Bouncing droplets generate waves periodically. 
They would not be wave sources in very specific conditions, for example in a confined system, it would correspond to the case where the bouncing droplet is located at a node of the excited mode~\cite{yc_SO-eingenstates14}.}.

It is easy to retrieve the previous results about the free bead in the low velocity approximation by applying the equivalent Schrödinger equation (\ref{esim_schroeq}) on $\psi$. 
In this case, $\psi$ can be expressed as $F \; \e^{\ii\,(k\,x - \omega\,t)}$ where $F=F(\sqrt{(x-v\,t)^2 + y^2 + z^2})$~\footnote{The imaginary part coming from (\ref{esim_schroeq}) yields $\frac{\partial F}{\partial t}=-\frac{c_m^2}{\Omega_m}\frac{\partial F}{\partial x}$. 
As here $\frac{\partial F}{\partial t}=-v\frac{\partial F}{\partial x}$, it comes that $k=\frac{\Omega_m\,v}{c_m^2}$. 
The real part yields $\omega\,F=-2\frac{c_m^2}{\Omega_m}\,(\Delta F - k^2\,F)$, which is in agreement with $\Delta F = - k^2\,F$ and Eq.~(\ref{eth_omega}).}.
In the case of the free bead, we have already extensively discussed the correspondence between the kinetic of the bead and the characteristics of the wave. 
Now, we would like to extend it to systems which are commonly considered in quantum mechanics handbooks.

%%%%%%%%%%%%%
%%%%%%%%%%%%%
%%%%%%%%%%%%%

\subsection{Bead in a linear cavity}

We are looking for the modulating wave $\psi$ associated to the bead in symbiosis in a linear cavity of length $L$, when the wave is standing in the laboratory reference frame. 
We consider the case of boundary conditions such that the wave amplitude is null in $x=0$ and $x=L$.

We draw from Eq.~(\ref{esim_schroeq}) and the boundary conditions that $\psi(x,\,t) = A\, \sin(K_n\,x)\,\e^{-\ii\,\omega_n\,t}$, where $K_n = \frac{n\,\pi}{L}$ ($n$ being a natural number), $\omega_n = \frac{K_n^2\,c_m^2}{2\,\Omega_m}$ and $A$ denotes the maximum amplitude of $\psi$. 
The corresponding wave $\varphi$ (cf. Eq.~(\ref{eth_psi_def})) is also a standing wave and writes as $\varphi(x,\,t) = A\, \sin(K_n\,x)\,\cos((\Omega_m+\omega_n)\,t)$. 
But not every stationary solution given above is suitable, as the symbiosis condition between the wave and the bead must be satisfied.
Since $\varphi$ is standing in the laboratory reference frame, it is also the reference frame proper to the bead. 
Thus, the symbiosis condition (\ref{eth_symbiose}) implies that $\Omega_m + \omega_n = \Omega_0$. 
This leads to only one possible solution for fixed pulsations $\Omega_m$ and $\Omega_0$. 
We discuss this point of divergence with quantum mechanics in the conclusion section. \\

Let us interpret the previous results by means of the effective kinetic energy. 
We first evaluate the local wave-vector $\vec{\kappa}$, at the location of the bead. 
We can see here that $\vec{\kappa}$ is uniform and equal to $\pm\, K_n\, \vec{e}_x$. 
According to relations (\ref{eth_ve}), wherever it is located, the bead behaves as if it has an effective velocity with regards to the elastic medium such that
\be \label{esim_ve_cavlin}
\vecve = \pm K_n \, \frac{c_m^2}{\Omega_m}\,\vec{e}_x.
\ee
The sign $\pm$ denotes here not ``or else" but ``both". 
This reflects the superposition of states from the elastic medium point of view. 
In other words, from the elastic medium point of view, the bead is perceived as if it has simultaneously the effective velocity $\vecve=K_n \, \frac{c_m^2}{\Omega_m}\,\vec{e}_x$ and $\vecve=-K_n \, \frac{c_m^2}{\Omega_m}\,\vec{e}_x$ when the wave  $\varphi$ is standing and in symbiosis with the bead. 
Note that what an observer measures, that is here either a bead at rest or with transverse oscillations at a fixed point, is not what perceives  -- and acts on -- the elastic medium.

We can now retrieve usual results of the equivalent quantum system. 
We can express the effective kinetic energy of the bead ($\ece=\frac{1}{2}m_0\,\ve^2$) using Eq.~(\ref{esim_ve_cavlin}), $K_n = \frac{n\,\pi}{L}$ and the proportionality coefficient (\ref{eth_hbe}):
\be 
\ece = \frac{n^2\pi^2\hbe^2}{2m_0L^2} \,.
\ee
This form is analogous to the quantum energy $E_n$, of the equivalent quantum system (cf. {\it e.g.}~\cite{cohentannoudji,MQ-X}).

We have seen that the energy of a standing wave in a linear cavity is also equal to the kinetic energy of the bead.
In quantum mechanics, it is usually attributed to the wave $\psi$ via $-\ii\,\hbar\,\frac{\partial \psi}{\partial t}$.
Note that the kinetic of the bead is considered from the elastic medium point of view rather than from an observer, who would be connected to the laboratory reference frame.

However, the linear cavity does not allow us to investigate the correspondence between the  wave and the particle characteristics related to the angular momentum.
In order to scrutinize this aspect on an example, we consider a bead without any external force in a spherical cavity.

%%%%%%%%%%%%%
%%%%%%%%%%%%%
%%%%%%%%%%%%%

\subsection{Bead in a spherical cavity}

We are looking for the modulating wave $\psi$ associated to the bead in symbiosis in a spherical cavity of radius $R$, when the wave is standing in the laboratory reference frame. 
We assume that the field is null at the boundaries of the cavity.

Expressing the modulating wave with spherical coordinates and separating the variables, it comes that $\psi(r,\,\theta,\,\phi,\,t)=F(r)G(\theta)H(\phi)\,\e^{-\ii\,\omega\,t}$. 
The equivalent Schrödinger equation (\ref{esim_schroeq}) implies that $\Delta\psi=-K^2\,\psi$, where $\omega = \frac{K^2\,c_m^2}{2\,\Omega_m}$.
Expressing the Laplacian in spherical coordinates, a standard calculation (used for example in acoustics~\cite{acoustique} \S 6) yields: $\psi = A\:\mathrm{j}_\ell(Kr)\:P_\ell^{m}(\cos \theta)\,\cos(m\,\phi)\,\e^{-\ii\,\omega\,t}$, where $\mathrm{j}_\ell$ denotes the spherical Bessel function of the first kind and order $\ell$, $P_\ell^m$ is the associated Legendre polynomial (where $\ell$ and $m$ are two natural numbers such that $ |m| \leq \ell$) and $A$ the amplitude -- which has no substantial consequence in this section either.
Note that $\psi$ could be expressed with real spherical harmonics but not with complex ones, where $Y_\ell^m(\theta,\,\phi)\propto P_\ell^m(\cos\theta)\,\e^{\ii m\phi}$.
It comes from the fact that they would correspond to non-standing waves in the laboratory reference frame, due to the term $\e^{\ii (m\phi-\omega t)}$.
The boundary conditions imply that $\psi=0$ at $r=R$, thus $K$ is restricted to particular values $K_{n,\ell}$ such that $K_{n,\ell}R$ is equal to the $n$-th zero of $\mathrm{j}_\ell(X)$.
We obtain an expression equivalent to a quantum particle in a spherical cavity (see {\it e.g.}~\cite{landau_MQ} \S 33), except for the fact that spherical harmonics are complex in the quantum case.
So, the standing wave in the spherical cavity writes as $\varphi = A\:\mathrm{j}_\ell(K_{n,\ell}\,r)\:P_\ell^{m}(\cos \theta)\,\cos(m\,\phi)\,\cos((\Omega_m+\omega_{n,\ell})\,t)$, where boundary conditions are verified and $\omega_{n,\ell} = \frac{K_{n,\ell}^2\,c_m^2}{2\,\Omega_m}$. 
Besides that, we have to take into account the symbiosis condition (\ref{eth_symbiose}) between the bead and the wave, as in the case of linear cavities.
Again, it involves  that $\Omega_0 = \Omega_m + \omega_{n,\ell}$. 
With fixed pulsations $\Omega_m$ and $\Omega_0$, the symbiosis condition restricts all possible solutions to only one with the additional pulsation $\omega_{n,\ell}$, similarly to the linear cavities. 
Note that this solution is degenerate since it allows the existence or the superposition of several solutions with different $m$ for the same $\omega_{n,\ell}$.\\

Now, we investigate whether the effective velocity of the bead accounts for the wave characteristics associated to the system, in particular the energy, the magnitude of the angular momentum and the component along $(Oz)$ axis of this momentum. 
First, we evaluate the local wave-vector $\vec{\kappa}_M$ at the location of the bead $\vec{r}_M$, the index $M$ designates the location of the bead.
Expressing the Laplacian of $\psi$ with spherical coordinates as above, and by identification with $-(\kappa_{M,\,r}^2+\kappa_{M,\,\theta}^2+\kappa_{M,\,\phi}^2)\psi$, it comes that
\footnote{With $\psi=A\:\mathrm{j}_\ell(K_{n,\ell}\,r)\:P_\ell^{m}(\cos \theta)\,\cos(m\,\phi)\,\e^{-\ii\,\omega\,t}$ it comes that: 
$\frac{1}{r^2}\frac{\partial}{\partial r}(r^2 \frac{\partial \psi}{\partial r}) = - K_{n,\ell}^2 \psi + \frac{\ell\,(\ell+1)}{r^2}\psi$,
$\frac{1}{r^2 \sin \theta}\frac{\partial}{\partial \theta}(\sin \theta \frac{\partial \psi}{\partial \theta})
\\ = - \frac{\ell\,(\ell+1)}{r^2}\psi + \frac{m^2}{r^2\,\sin^2\theta}\psi$
and $\frac{1}{r^2 \sin^2 \theta}\frac{\partial^2 \psi}{\partial \phi^2} = -\frac{m^2}{r^2 \sin^2 \theta}\psi$.} 
$\kappa_{M,\,r}^2 = K_{n,\ell}^2 - \frac{\ell\,(\ell+1)}{r_M^2}$, 
$\kappa_{M,\,\theta}^2 = \frac{\ell\,(\ell+1)}{r_M^2} - \frac{m^2}{r_M^2\,\sin^2\theta_M}$ 
and $\kappa_{M,\,\phi}^2 = \frac{m^2}{r_M^2\,\sin^2\theta_M}$. 
According to the relation (\ref{eth_ve}), the bead behaves as if it has an effective velocity $\vecve$ with regards to the elastic medium, such that:
\be \label{esim_ve_cavsphe}
%\vecve = \frac{c_m^2}{\Omega_m}\, \coortri{\pm\, \sqrt{K^2 - \frac{\ell\,(\ell+1)}{r_M^2}}}{\pm \, \sqrt{\frac{\ell\,(\ell+1)}{r_M^2} - \frac{m^2}{r_M^2\,\sin^2\theta_M}}}{\pm \, \sqrt{\frac{m^2}{r_M^2\,\sin^2\theta_M}}}
v_{\mathrm{eff},\,r} = \pm \frac{c_m^2}{\Omega_m}\,\sqrt{K_{n,\ell}^2 - \frac{\ell\,(\ell+1)}{r_M^2}}\,,~v_{\mathrm{eff},\,\theta} = \pm \frac{c_m^2}{\Omega_m}\,\sqrt{\frac{\ell\,(\ell+1)}{r_M^2} - \frac{m^2}{r_M^2\,\sin^2\theta_M}}\, ~\mathrm{et}~ v_{\mathrm{eff},\,\phi} = \pm \frac{c_m^2}{\Omega_m}\,\sqrt{\frac{m^2}{r_M^2\,\sin^2\theta_M}}\,.
\ee

We can evaluate the effective angular momentum of the bead defined as 
\be \vecLe = m_0\, \vec{r}_M  \times \vecve \,.
\ee
Thus, $\vecLe = -(m_0\,r_M\,v_{\mathrm{eff},\,\phi})\vec{e}_{\theta} + (m_0\,r_M\,v_{\mathrm{eff},\,\theta})\vec{e}_{\phi}$, and its component along the $(Oz)$ axis is $L_{\mathrm{eff},\,z} = \vecLe\cdot\vec{e}_z=(m_0\,r_M\,v_{\mathrm{eff},\,\phi})\sin \theta$. 
Then, we can deduce its magnitude $\vecLe^2=m_0^2\,\frac{c_m^4}{\Omega_m^2}\,\ell\,(\ell+1)$, so its component along the $(Oz)$ axis can be expressed as $L_{\mathrm{eff},\,z}= \pm \, m_0\,\frac{c_m^2}{\Omega_m}\,m$. 
According to the expression of the proportionality coefficient (\ref{eth_hbe}), it comes that
\be 
\vecLe^2=\hbe^2\,\ell\,(\ell+1)\,, ~~\mathrm{and}~~ L_{\mathrm{eff},\,z}= \pm \,\hbe \,m \,.
\ee

Here again $\pm$ denotes ``both", as for Eq.~(\ref{esim_ve_cavlin}). 
In the laboratory reference frame, the wave $\varphi$ (and $\psi$) are standing and the bead is at rest ({\it i.e.} its velocity is null). 
This is in agreement with our intuition about the symbiosis between the wave and the bead. 
It explains why the component of the effective angular momentum along $(Oz)$ axis $L_{\mathrm{eff},\,z}$, is simultaneously $\hbe \,m$ and $-\hbe \,m$. 
A unique value would contradict the observation of a bead at rest in the laboratory frame. 
But it contrasts with quantum mechanics, according to which a stationary state with the expression $Y_\ell^m(\theta,\,\phi)\propto P_\ell^m(\cos\theta)\,\e^{\ii m\phi}$ for the $(\theta,\,\phi)$ coordinates has only one value of $L_{\mathrm{eff},\,z}$, that is $\hbar\, m$. 
However, as already mentioned, a wave $\varphi$ with complex spherical harmonics would not be standing in the laboratory frame. 
The expression of $\varphi$ above can be seen as the superposition of two functions, one with $Y_\ell^m(\theta,\,\phi)$ and the other one with $Y_\ell^{-m}(\theta,\,\phi)$. 
Thus, this is in agreement with the point under discussion, $L_{\mathrm{eff},\,z}$ is both $\hbe \,m$ and $-\hbe \,m$. 
Then, the total component along $(Oz)$ axis of the effective angular momentum is null, which is in agreement with a bead at rest in the laboratory frame -- as in the case of the previous example of a linear cavity, where the total effective linear momentum is also null.

The effective kinetic energy of the bead, $\ece=\frac{1}{2}m_0\,\ve^2$ ($=\frac{1}{2}m_0\,v_{\mathrm{eff},\,r}^2 + \frac{\Le^2}{2\,m_0\,r_M^2}$ here) is, according to Eqs.~(\ref{esim_ve_cavsphe}) and (\ref{eth_hbe}), $\ece=\frac{\hbe^2\,K_{n,\ell}^2}{2\,m_0}$. 
Note that the location of the bead has no impact on the values of $\ece$, $\Le^2$ and $L_{\mathrm{eff},\,z}$.
We obtain values of $\ece$, $\Le^2$ and $L_{\mathrm{eff},\,z}$ which are equivalent to the values of these quantities, associated to the wave, in the equivalent quantum mechanics system.

To summarize the case of a spherical cavity, when there is symbiosis between the bead and the wave and the wave is standing in the laboratory reference frame, the wave characteristics correspond to the particle ones.
Note still that these characteristics are established from the effective kinetic of the bead and not from what an observer sees. 
The effective kinetic of the bead contains the information necessary to compute the characteristics which are commonly assigned to the wave in quantum mechanics. 
Thus, there is a concordance between the wave and the particle representation, when they account for the state of the system -- in particular the energy, the magnitude and the component along an axis of the angular momentum.

Lastly, we can be surprised by how simple the particle-like representation is, in comparison to the one commonly encountered in quantum mechanics handbooks (cf. {\it e.g.}~\cite{cohentannoudji,MQ-X,landau_MQ}).

%%%%%%%%%%%%%%%%%%%%%%%%%%
%%%%%%%%%%%%%%%%%%%%%%%%%%
%%%%%%%%%%%%%%%%%%%%%%%%%%

\section*{Conclusion}

In this paper, we suggested a dual wave-particle system at macroscopic scale.
It is based on two series of experiments, the main elements are drawn from the bouncing droplets experiments (see~\cite{bush_review15} for a review), while the simple mathematical formalisation stems from the sliding bead along an oscillating string experiment~\cite{yc_sao1999}.
The system consists in (1) a bead oscillator, {\it i.e.} a punctual mass with an ``internal clock'' which forces the mass to oscillate at a proper pulsation $\Omega_0$ through a quadratic potential, and (2) an elastic medium which carries transverse waves $\varphi$, governed at this step by the d'Alembert equation with wave speed $c_m$.
In addition, the medium is prone to support standing vibrations at pulsation $\Omega_m$ in the laboratory reference frame, through a quadratic potential. 
Waves in the medium are thus governed by a Klein-Gordon like equation. 
The bead generates transverse waves in the elastic medium and reciprocally the transverse waves guide the bead. 
This remind us of the ``pilot-wave" suggested by de Broglie and observed for the first time in a hydrodynamic framework through the bouncing droplets experiments.

The coupled equations of the motion of the bead~(\ref{eth_mvt}) and of the waves~(\ref{eth_ch}) are established by using a Lagrangian formalism. 
It is important to notice that calculations are covariant according to the elastic medium, {\it i.e.} the Lorentz-Poincaré transformation is expressed with $c_m$ in order to respect the invariance of the d'Alembert or Klein-Gordon equation of the medium.

We have considered the special case where the bead and the wave are in intimate harmony, called here symbiosis. 
In this situation, the bead is neither the source of the wave nor is it deflected by the wave. 
The bead oscillates at its natural pulsation $\Omega_0$ in its proper reference frame and its energy remains constant.
Once this state is reached, the bead is in the wave field and yet behaves as if it was not. 
The expression of the wave for a free bead has been evaluated~(\ref{eth_varphiR}). 
More importantly, we have introduce (1) the modulating wave $\psi$ (\ref{eth_psi_def}) which designates the wave due to the bead in contrast to the natural wave of the elastic medium and (2) the effective velocity of the bead with regards to the elastic medium.
The latter is defined~(\ref{eth_ve}) as the velocity of the bead from the point of view of the elastic medium.
The effective kinetic energy and momentum are then proportional to the wave characteristics, more precisely the pulsation and the local wave-vector. 
The proportionality factor~(\ref{eth_hbe}) which allows to make a correspondence between particle and wave characteristics, is not proper to the elastic medium, as it also depends on the mass of the bead -- which is arbitrary in the suggested system.

In the state of symbiosis, without external action and under the low velocity approximation, we have shown that $\psi$ is governed by an equation which is strictly equivalent to Schrödinger equation.
When the bead in symbiosis is in a linear or a spherical cavity and when the wave $\varphi$ is standing in the laboratory reference frame, we have confirmed in our system what de Broglie had suggested, that is to say that the (effective) kinetic of the bead accounts for quantities which are commonly attributed to the wave-like nature of the system in quantum mechanics. 
More precisely, the magnitude, a component of the angular momentum and the energy of the system, which are evaluated from $\psi$ in quantum mechanics, are in exact agreement with the effective kinetic of the bead. 
Note also that these values can be simply deduced from the effective kinetic of the bead.

The wave $\varphi$ does not exhibit any singularity. 
In the context of our system, this validates  the ``pilot-wave" hypothesis compared to the ``double solution" hypothesis suggested by de Broglie~\cite{ldb_tentative,ldb_interpretation1959}. 
Moreover, the wave $\varphi$ (and then $\psi$) has a \textit{real} nature -- and not a statistical one -- in the sense that it accounts for the field of transverse waves in the system.

In this paper, we pursued the following goal: to devise a system which would feature the main characteristics of the bouncing droplets experiments, while maintaining a convenient formalism to identify quantum analogies (and differences).
In order to expand this investigation toward deeper quantum analogies, we should probably go over the symbiosis condition, because it restrains  all the possible stationary solutions to only one, as for example in the case of cavities. 
Moreover, it could also be interesting to study the bead without its ``internal clock'', it would mean that $\Omega_0=0$, which {\it de facto} eliminates the symbiosis condition.
This would bring the system a little closer to the bouncing droplets experiments.

%%%%%%%%%%%%%
%%%%%%%%%%%%%
%%%%%%%%%%%%%

\subsubsection*{Aknowledgments}

We thank Hervé Mohrbach, Alain Bérard, Laurent Bindel, Yves Couder, Navid Nemati and Alexandra for discussions and support. A great thank to Lionel Tabourier notably for his great support with this translation.

%%%%%%%%%%%%%%%%%%%%%%%%%%
%%%%%%%%%%%%%%%%%%%%%%%%%%
%%%%%%%%%%%%%%%%%%%%%%%%%%
%\appendix 

\section*{Appendices}

\renewcommand{\theequation}{A\arabic{equation}}% redefine the command that creates the equation no.
\setcounter{equation}{0}  % reset counter 
\renewcommand{\thesubsection}{A1}

\subsection{Calculation of the wave equation~(\ref{eexpe_ch}) \label{sa_expe_ech}}

In this appendix, we only consider the generalised Euler-Lagrange equation ($\frac{\partial (\mathcal{L}_{f} + \mathcal{L}_{i} + \mathcal{L}_{e})}{\partial \varphi} = \frac{\partial}{\partial t}\frac{\partial (\mathcal{L}_{f} + \mathcal{L}_{i} + \mathcal{L}_{e})}{\partial (\partial \varphi/\partial t)} + \frac{\partial}{\partial x}\frac{\partial (\mathcal{L}_{f} + \mathcal{L}_{i} + \mathcal{L}_{e})}{\partial (\partial \varphi/\partial x)}$) related to the interaction Lagrangian density, $\mathcal{L}_{i}$. 
Calculations of the Lagrangian densities of the field and of the external excitation are common in the literature.

The interaction Lagrangian density writes as $\mathcal{L}_i = \frac{m_0}{2}\, \left(\frac{\dd \varphi(x,t)}{\dd t}\right)^2 \cdot \delta\left(x - \xi(t)\right)$, where $\frac{\dd \varphi}{\dd t} = \frac{\partial \varphi}{\partial t} + \vec{v}\cdot\vec{\nabla}\varphi$ and $\delta$ denotes the Dirac distribution. 
It comes that $\frac{\partial \mathcal{L}_{i}}{\partial \varphi} = 0$,
\be
\frac{\partial}{\partial t}\frac{\partial \mathcal{L}_{i}}{\partial (\partial \varphi/\partial t)} = \frac{\partial}{\partial t}\left(m_0\,\frac{\dd \varphi}{\dd t}\:\delta\left(x - \xi(t)\right) \right) = m_0\,\frac{\partial}{\partial t}\left(\frac{\dd \varphi}{\dd t}\right)\:\delta\left(x - \xi(t)\right)\: -\: m_0\,\frac{\dd \varphi}{\dd t}\, \vec{v}\cdot\vec{\nabla}\left(\delta(x - \xi(t))\right)
\ee
and
\be  
\frac{\partial}{\partial x}\frac{\partial \mathcal{L}_{i}}{\partial (\partial \varphi/\partial x)} =
\frac{\partial}{\partial x}\left(m_0\,v_x\,\frac{\dd \varphi}{\dd t}\:\delta\left(x - \xi(t)\right) \right) =
m_0\, v_x\,\frac{\partial}{\partial x}\left(\frac{\dd \varphi}{\dd t}\right) \: \delta\left(x - \xi(t)\right)\: + \: m_0\,\frac{\dd \varphi}{\dd t}\,v_x\,\frac{\partial}{\partial x}\left(\delta\left(x - \xi(t)\right)\right).
\ee
So, the sum of the two previous equations leads to
\be 
 m_0\, \left[\frac{\partial}{\partial t}\left(\frac{\dd \varphi}{\dd t}\right) + \vec{v}\cdot\vec{\nabla}\left(\frac{\dd \varphi}{\dd t}\right)\right] \:\delta\left(x - \xi(t)\right) = \,m_0 \frac{\dd^2 \varphi}{\dd t^2} \: \delta\left(x - \xi(t)\right) \,,
\ee
where the particle acceleration of the bead appears.

%%%%%%%%%%%%%
%%%%%%%%%%%%%
%%%%%%%%%%%%%

\renewcommand{\thesubsection}{A2}

\subsection{Calculation of the motion equation~(\ref{eth_mvt}) \label{sa_th_emvt}}

The motion equation of the bead comes from a principle of least action, when the four-position of the bead is subjected to a small change, $\xi^\mu \rightarrow \xi^\mu + \delta\xi^\mu$, while the field $\varphi$ is fixed.
Such calculations are usual in analytical mechanics with relativistic formulation.
$L_m$ is the only Lagrangian here which depends on the location of the bead. 
Thus, we are looking for the conditions for which the small change $\delta\xi^\mu$ implies at the first order $\delta(\int L_m\, \dd \tau) = 0$, where boundary values are fixed.

With $\xi^\mu \rightarrow \xi^\mu + \delta\xi^\mu$, it comes $\varphi(\xi^\mu) \rightarrow \varphi(\xi^\mu) + \partial_\mu \varphi \: \delta\xi^\mu$ et $\dd \tau = \sqrt{\dd \xi^\mu \dd \xi_\mu}/c \rightarrow \dd \tau + \delta(\dd \tau)$ where $\delta(\dd \tau) = \frac{U_\mu}{c_m^2}\, \dd(\delta\xi^\mu)$, and $U^\mu=\frac{\dd \xi^\mu}{\dd \tau}$.

The small change of the Lagrangian of the bead in its proper reference frame (cf. Eq.~(\ref{eth_Lm} where $\gamma=1$) writes as
\be 
\delta(L_m) = m_0\,\left[\frac{\dd \varphi}{\dd \tau} \frac{\dd(\partial_\mu \varphi \: \delta\xi^\mu)}{\dd \tau} \; - \; \left(\frac{\dd \varphi}{\dd \tau}\right)^2\,\frac{\delta(\dd \tau)}{\dd \tau} \; - \;\frac{\Omega_0^2}{c_m^2}\,\varphi\, \partial_\mu \varphi \: \delta\xi^\mu \right] \,.
\ee

$\delta(\int L_m\, \dd \tau) = 0$ implies $\int \delta(L_m)\, \dd \tau + \int L_m \, \frac{\delta(\dd \tau)}{\dd \tau}\, \dd \tau = 0$. So it comes that
\be 
m_0 \int \frac{\dd \varphi}{\dd \tau} \frac{\dd(\partial_\mu \varphi \: \delta\xi^\mu)}{\dd \tau} \dd\tau \; - \;\frac{m_0}{2}\int \left(\left(\frac{\dd \varphi}{\dd \tau}\right)^2 + \frac{\Omega_0^2}{c_m^2}\,\varphi^2 \right)\, \frac{\delta(\dd \tau)}{\dd \tau} \, \dd \tau \; - \; m_0\int \frac{\Omega_0^2}{c_m^2}\,\varphi\, \partial_\mu \varphi \: \delta\xi^\mu \dd \tau = 0 \,.
\ee
Integrating by parts the two integrals on the left side, with fixed end points, the previous expression of $\delta(\dd \tau)$ leads to
\be 
-\ m_0 \int \frac{\dd^2 \varphi}{\dd \tau^2} \partial_\mu \varphi \; \delta\xi^\mu \dd \tau \; + \; \frac{m_0}{2}\int \frac{\dd}{\dd \tau}\left[\left(\left(\frac{\dd \varphi}{\dd \tau}\right)^2 + \frac{\Omega_0^2}{c_m^2}\,\varphi^2 \right)\, \frac{U_\mu}{c_m^2}\right] \: \delta\xi^\mu \dd \tau \; - \; m_0\int \frac{\Omega_0^2}{c_m^2}\,\varphi\, \partial_\mu \varphi \; \delta\xi^\mu \dd \tau = 0 \,.
\ee
Since the small change $\delta\xi^\mu$ is arbitrary, the motion equation (\ref{eth_mvt}) follows.

Note that the Euler-Lagrange equation, $\frac{\dd}{\dd t}\frac{\partial L_m}{\partial \vec{v}} = \frac{\partial L_m}{\partial \vec{\xi}}$, where $L_m$ is formulated in Eq.~(\ref{eth_Lm}), provides the motion equation~(\ref{eth_mvt_noncov}).

%%%%%%%%%%%%%
%%%%%%%%%%%%%
%%%%%%%%%%%%%

\renewcommand{\thesubsection}{A3}

\subsection{Calculation of the wave equation~(\ref{eth_ch}) \label{sa_th_ech}}

The wave equation comes from a principle of least action, when the field is subjected to a small change, $\varphi \rightarrow \varphi + \delta\varphi$, while the four-position of the bead is fixed.
We are thus looking for the conditions for which the small change $\delta \varphi$ implies at first order $\delta \int (\mathcal{L}_m + \mathcal{L}_{f}) \dd t \dd^3 \vec{r} = 0$, where boundary values are fixed.

With $\varphi \rightarrow \varphi + \delta\varphi$, it comes that $\delta(\frac{\dd \varphi}{\dd \tau})^2 = 2\frac{\dd \varphi}{\dd \tau}\frac{\dd (\delta \varphi)}{\dd \tau}$ and $\delta(\partial_\mu \varphi\,\partial^\mu \varphi)=2\partial_\mu (\delta\varphi)\,\partial^\mu \varphi$.

According to the Lagrangian densities given in Eqs.~(\ref{eth_lm}) and (\ref{eth_lch}), it comes that
\be 
\int \left[ \rho_0 \left(\frac{\dd \varphi}{\dd \tau}\frac{\dd (\delta \varphi)}{\dd \tau} - \frac{\Omega_0^2}{c_m^2}\varphi\:\delta \varphi\right) + \mathcal{T}\,\partial_\mu (\delta\varphi)\,\partial^\mu \varphi - \mathcal{T}\,\frac{\Omega_m^2}{c_m^2}\varphi\:\delta \varphi \right] \dd t \dd^3 \vec{r} = 0 \,.
\ee
Integrating by parts the terms  $\frac{\dd (\delta \varphi)}{\dd \tau}$ and $\partial_\mu (\delta\varphi)$, with fixed end points, provides
\be
\int \left[-\,\frac{\dd}{\dd \tau}\left(\rho_0\frac{\dd \varphi}{\dd \tau}\right) - \rho_0\frac{\Omega_0^2}{c_m^2}\varphi - \mathcal{T}\,\partial_\mu\partial^\mu\varphi - \mathcal{T}\,\frac{\Omega_m^2}{c_m^2}\varphi \right]\,\delta \varphi \; \dd t \dd^3 \vec{r} = 0 \,.
\ee
Since the small change $\delta\varphi$ is arbitrary and according to the law of conservation of the mass ($\frac{\dd \rho_0}{\dd \tau} = 0$), the wave equation~(\ref{eth_ch}) follows.

Note that the generalised Euler-Lagrange equation and the conservation of the mass provide the same result, as expected.

%%%%%%%%%%%%%%%%%%%%%%%%%%
%%%%%%%%%%%%%%%%%%%%%%%%%%
%%%%%%%%%%%%%%%%%%%%%%%%%%

\end{document}